\newcommand{\be}{\begin{equation}}
\newcommand{\ee}{\begin{equation}}
\newcommand{\bea}{\begin{eqnarray}}
\newcommand{\bee}{\end{eqnarray}}
\newcommand{\bq}{\begin{quote}}
\newcommand{\eq}{\end{quote}}

\newcommand{\vecvar}[1]{\mbox{\boldmath$#1$}}

\newcommand{\degree}{\kern-.2em\r{}\kern-.3em}
\newcommand{\degC}{\kern-.2em\r{}\kern-.3em C}
\newcommand{\eref}[1]{Eq.~(\ref{#1})}

\newcommand{\fref}[1]{Fig.~\ref{#1}}

\newcommand{\sref}[1]{Sec.~\ref{#1}}

\documentclass[preprint,showpacs,preprintnumbers,amsmath,amssymb]{revtex4}

\usepackage{graphicx}% Include figure files
\usepackage{dcolumn}% Align table columns on decimal point
\usepackage{bm}% bold math

\begin{document}

%\pdfbookmark

%\preprint{APS/123-QED}

\title
{%
Distribution of the spacing between two adjacent avoided crossings
}

\author{
Manabu Machida$^{1*}$ and Keiji Saito$^{2\dag}$
}

\affiliation{%
$^1$Institute of Industrial Science, The University of Tokyo, 
Meguro-ku, Tokyo 153-8505, Japan\\
$^2$Department of Physics, Graduate School of Science, 
The University of Tokyo, Bunkyo-ku, Tokyo 113-0033, Japan 
}

\date{\today}

\begin{abstract}
We consider the frequency at which avoided crossings appear in an energy 
level structure when an external field is applied to a quantum 
chaotic system.  The distribution of the spacing in the parameter 
between two adjacent avoided crossings is investigated.
Using a random matrix model, we find that
the distribution of these spacings is well fitted by a power-law 
distribution for small spacings.  
The powers are $2$ and $3$ for the Gaussian orthogonal 
ensemble and Gaussian unitary ensemble, respectively.  We also find that 
the distributions decay exponentially for large spacings. The 
distributions in concrete quantum chaotic systems agree 
with those of the random matrix model.  
\end{abstract}



\maketitle

%%%%%%%%%%%%%%%%%%%%%%%%%%%%%%%%%%%%%%%%%%%%%%%%%%%%%%%%%%%%%%%%%%%
\section{Introduction}
The avoided crossing is a ubiquitous structure in energy 
levels of quantum chaotic systems with external perturbation
\cite{Haake}.  
As the time-dependent external perturbation is applied to the system,
the quantum state changes nonadiabatically.
Especially when the perturbation changes linearly and slowly in time, 
nonadiabatic transitions can be described by 
the Landau-Zener transitions \cite{lz} at avoided crossings.  
Using this microscopic mechanism, 
the nonadiabatic change of the total quantum state might be understood. 
Energy diffusion phenomena have been intensively studied 
from this point of view \cite{Gefen87,Wilkinson88,Wilkinson90,Machida02}.  
To understand macroscopic phenomena such as energy diffusion,
it is important to study universal aspects of avoided crossings in
quantum systems. Thus, using random matrices\cite{Mehta,Porter}, 
various distributions concerning avoided crossings have been studied
\cite{Guhr98}.  
For example, the distributions of the energy level curvature
\cite{Gaspard89,Gaspard90,Zakrzewski93b,vonOppen94,vonOppen95}, 
the difference between the slopes of the asymptotic lines
\cite{Wilkinson89,Zakrzewski93a}, and 
the minimum energy gap \cite{Zakrzewski93a,Zakrzewski91} are derived 
at avoided crossings, 
and their universalities are confirmed in concrete quantum 
systems \cite{Gaspard90,Zakrzewski93b,Zakrzewski93a,Zakrzewski91,TH92}.  

In this paper, we consider another distribution related to 
the structure of avoided crossings.
We study the distribution of the spacing 
in a parameter space of perturbation, which is measured as the distance 
between two adjacent avoided crossings involved in two neighboring 
energy levels (examples of this spacing are depicted in \fref{levels}.)  
We call this distribution the avoided-crossing spacing distribution (ACSD).  
Only a few qualitative studies related to the ACSD have been published.  
Goldberg and Schweizer obtained the ACSD of the hydrogen atom in the 
magnetic field and that of the Africa billiard to estimate the 
statistical error of the gap distribution \cite{Goldberg91}.  
Wilkinson and Austin discussed the density of avoided crossings
in quantum systems with two free parameters, although the distribution 
of the spacing between avoided crossings was not 
discussed \cite{Wilkinson93}.  
The main aim of this paper is to obtain the {\em quantitative} properties of 
the ACSD.

We first consider the Hamiltonian taken from random matrices 
to extract the general features of the ACSD. 
We calculate the ACSDs for the Gaussian orthogonal ensemble (GOE) 
and the Gaussian unitary ensemble (GUE) by a precise numerical 
calculation.  We find that, for small spacings, the ACSD for GOE (GUE) 
shows the power law behavior with the power $2$ ($3$).  For large 
spacings, the ACSDs decay exponentially.  We then consider two 
concrete quantum systems, i.e., the coupled rotators model and the 
Aharonov-Bohm billiard, and find that the ACSDs of these systems agree 
with those predicted by simulations using random matrices.  

This paper is constructed as follows.  In \sref{rm}, we obtain the 
ACSDs of a random matrix model for GOE and GUE, and discuss 
differences between them, focusing especially on the powers of the 
distributions.  
In \sref{crt} and \sref{ab}, we present the ACSDs of two concrete 
models and show that they are identical to those of the corresponding 
random matrix models.  Finally, in \sref{disc}, 
we give a summary.  

%%%%%%%%%%%%%%%%%%%%%%%%%%%%%%%%%%%%%%%%%%%%%%%%%%%%%%%%%%%%%%%%%%%
\section{Random Matrix Model}
\label{rm}

Let us consider an $N\times N$ Hamiltonian matrix 
$\mathcal{H}(\Lambda)$ including a parameter $\Lambda$ as 
an external field.  
In order to capture general features of the ACSD, we take 
$\mathcal{H}(\Lambda)$ from random matrix ensembles \cite{Haake,Mehta}.  
We prepare random matrices $\mathcal{H}_0^{\rm RM}$ and 
$\mathcal{V}^{\rm RM}$ of $N=1000$.  
We consider the following random matrix model.  
\begin{equation}
 \mathcal{H}_{\rm RM}(\Lambda) = 
 \mathcal{H}_0^{\rm RM} \cos\Lambda + 
 \mathcal{V}^{\rm RM} \sin\Lambda.
\label{rmmodel1}
\end{equation}
Note that this model keeps the variance of matrix elements 
unchanged as the parameter varies \cite{Zakrzewski93a}.  

If we take both $\mathcal{H}_0^{\rm RM}$ and $\mathcal{V}^{\rm RM}$ 
from GOE, the Hamiltonian $\mathcal{H}_{\rm RM}(\Lambda)$ represents
quantum systems invariant under an antiunitary transformation such as 
the time-reversal transformation.  
On the other hand, if we take both of them from GUE, 
the Hamiltonian represents quantum systems that have no symmetries for 
antiunitary transformations \cite{Haake,Porter,Bohigas84}.  
We let $\Lambda$ move from $0.5$ to $1.5$.  
The variance $\sigma_{\rm RM}^2$ of diagonal elements of 
$\mathcal{H}_0^{\rm RM}$ and $\mathcal{V}^{\rm RM}$ is $2$.  
  
\begin{figure}[h]
\begin{center}
\includegraphics[scale=1.0,clip]{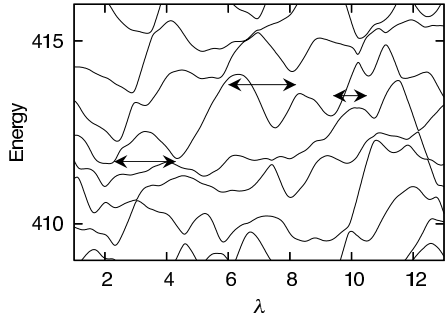}
\end{center}
\caption{The scaled energy spectrum for $\mathcal{H}_{\rm RM}(\Lambda)$ 
of GOE as a function of $\lambda$.  
Three examples of avoided-crossing spacings are also shown.}
\label{levels}
\end{figure}

First of all, we need to scale $\Lambda$ and $E_j(\Lambda)$ 
(the $j$-th eigenvalue of the instantaneous Hamiltonian 
$\mathcal{H}(\Lambda)$) in order to eliminate the dependence of 
the parameters on the external field and energy
\cite{Simons93a,Simons93b,Simons93c,Alhassid95a,Attias95a}.  
Following Simons and Altshuler \cite{Simons93a}, we define 
scaled parameter $\lambda$ and scaled energy $\epsilon_j$ as 
\begin{eqnarray}
\epsilon_j(\Lambda) 
&=& \epsilon_1(\Lambda) + \sum_{i=1}^{j-1}
\frac{E_{i+1}(\Lambda)-E_i(\Lambda)}{\Delta_i(\Lambda)} \nonumber \\
\lambda_{kj} 
&=& \lambda_{0j} + \sum_{k'=1}^{k-1}
(\Lambda_{k'+1}-\Lambda_{k'})\sqrt{\left\langle\left(
\frac{\partial\epsilon_j(\Lambda)}{\partial\Lambda}\biggr|_{\Lambda_{k'}}
\right)^2\right\rangle},\nonumber \\
&&
\label{scaling}
\end{eqnarray}
where $\Delta_j(\Lambda)$ is the mean level spacing at $E_j(\Lambda)$, 
and $\langle\cdots\rangle$ denotes a statistical average over 
a typical range of levels and $\Lambda$.  
We put $\epsilon_1(\Lambda)=0\,(=\mbox{const.})$, which is followed by 
$\langle\partial\epsilon_j(\Lambda)/\partial\Lambda\rangle=0$.  
$\lambda_{0j}$ is set to $\Lambda_0$.  
Thus, we obtain the scaled energy level flow $\epsilon_j(\lambda)$.  
We show the scaled energy spectrum $\epsilon_j(\lambda)$ for 
$\mathcal{H}_{\rm RM}(\Lambda)$ of GOE in \fref{levels}.  
Some avoided crossings look like crossings indeed, but 
this is due solely to the line width.  
In \fref{levels}, three distances (the right and left arrows) are shown as 
examples of avoided-crossing spacing.  

An avoided-crossing spacing is defined by the spacing between two 
neighboring avoided crossings on the same energy level flow.  
Now, $\lambda_{kj}$ is identified as the position of an avoided crossing 
on the $j$-th energy level flow if 
$\epsilon_{j+1}(\lambda_{k-1,j+1})-\epsilon_{j}(\lambda_{k-1,j})
>\epsilon_{j+1}(\lambda_{k,j+1})-\epsilon_{j}(\lambda_{k,j})$ and 
$\epsilon_{j+1}(\lambda_{k+1,j+1})-\epsilon_{j}(\lambda_{k+1,j})
>\epsilon_{j+1}(\lambda_{k,j+1})-\epsilon_{j}(\lambda_{k,j})$
are satisfied.  
By taking the differences in positions between two adjacent avoided 
crossings, avoided-crossing spacings $S_{\lambda}$ are obtained. 
We thereby obtain the ACSD $P(S_{\lambda})$, i.e., the probability 
that the spacing in the parameter 
space of the adjacent avoided crossings is $S_{\lambda}$.  Here 
$P(S_{\lambda})$ satisfies the following normalization conditions.  
\begin{equation}
 \int_0^{\infty}P(S_{\lambda}) {\rm d}S_{\lambda} = 1, \quad 
 \int_0^{\infty}S_{\lambda} P(S_{\lambda}) {\rm d}S_{\lambda} = 1.
\label{normalization}
\end{equation} 

Hereafter, we distinguish between GOE and GUE by superscript $\beta$.  
We have $\beta= 1$ for GOE and $\beta=2$ for GUE.  That is, 
$P_{\rm RM}^{(\beta)}(S_{\lambda})$ denotes the ACSD of model 
(\ref{rmmodel1}) for GOE and GUE.  
We take the parameter slice 
$\Delta\Lambda\equiv\Lambda_{k+1}-\Lambda_k$ 
to be $0.001$, which is small enough.  
We find $4.0\times10^5\,(3.5\times10^5)$ avoided-crossing spacings in 
50 samples of $\mathcal{H}_{\rm RM}(S_{\lambda})$ for GOE (GUE).  

\begin{figure*}[t]
\begin{center}
\includegraphics[scale=1.0,clip]{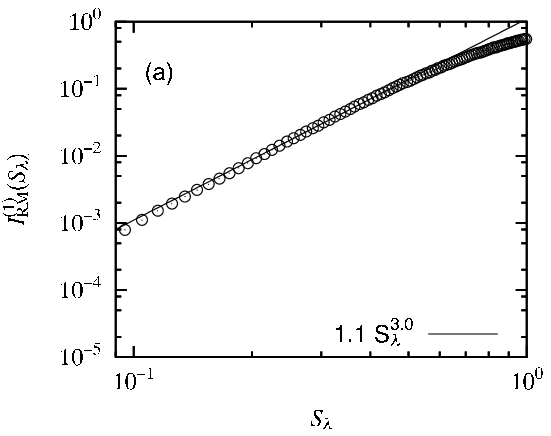}
\includegraphics[scale=1.0,clip]{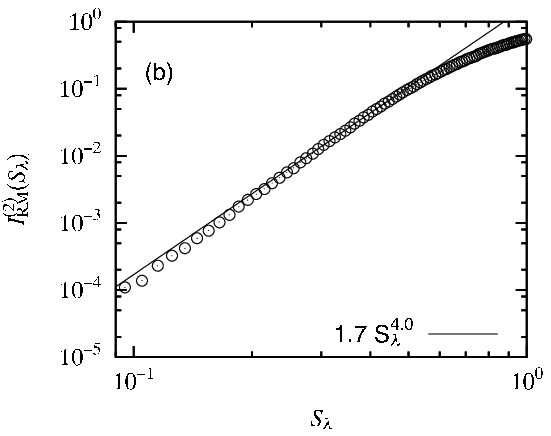}
\end{center}
\caption{
Log-log plots of the cumulative distributions 
(a) $I_{\rm RM}^{(1)}(S_{\lambda})$ and 
(b) $I_{\rm RM}^{(2)}(S_{\lambda})$.  
}
\label{int_tati}
\end{figure*}

We also define the cumulative distribution 
$I_{\rm RM}^{(\beta)}(S_{\lambda})$ of the ACSD of 
$\mathcal{H}_{\rm RM}(\lambda)$ for GOE and GUE as 
$I_{\rm RM}^{(\beta)}(S_{\lambda})=\int_{0}^{S_{\lambda}} 
P_{\rm RM}^{(\beta)}(S_{\lambda}') {\rm d}S_{\lambda}'$.  
Figures \ref{int_tati} show log-log plots of 
$I_{\rm RM}^{(\beta)}(S_{\lambda})$.  
The solid lines in the figures are obtained by fitting the data 
in the small $S_{\lambda}$ region by the least-squares method.  
Consequently, we find that $I_{\rm RM}^{(1)}(S_{\lambda})$ behaves as 
$1.1\times S_{\lambda}^{3.0}$ and that 
$I_{\rm RM}^{(2)}(S_{\lambda})$ behaves as 
$1.7\times S_{\lambda}^{4.0}$.  Therefore, we may write 
\begin{equation}
P_{\rm RM}^{(\beta)}(S_{\lambda}) \sim S_{\lambda}^{\beta+1} \quad 
(S_{\lambda}\ll 1).
\end{equation}

\begin{figure*}[t]
\begin{center}
\includegraphics[scale=1.0,clip]{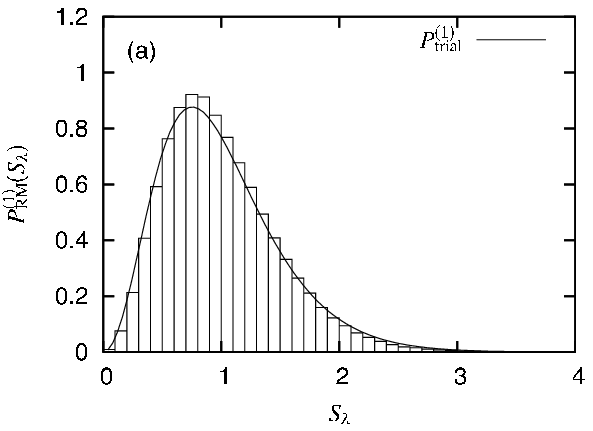}
\includegraphics[scale=1.0,clip]{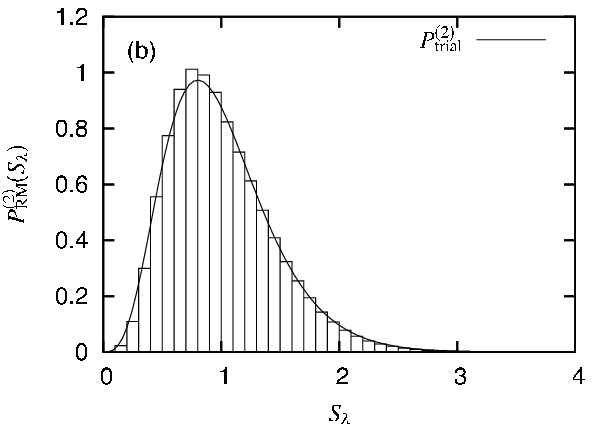}
\end{center}
\caption{
(a) For $\beta=1$, 
$P_{\rm RM}^{(\beta)}(S_{\lambda})$ is compared with 
$P_{\rm trial}^{(\beta)}(S_{\lambda})$.  
(b) The same as (a) except $\beta=2$.  
}
\label{trial}
\end{figure*}

To study how the ACSD decays, 
we compare $P_{\rm RM}^{(\beta)}(S_{\lambda})$ with the 
following trial function.  
\begin{equation}
 P^{(\beta)}_{\rm trial} (S_{\lambda}) = \frac{a_{\beta}S_{\lambda}^{\beta+2}}
 {\sinh{b_{\beta}S_{\lambda}}}.
\label{trialeq}
\end{equation}
Here, coefficients $a_{\beta}$ and $b_{\beta}$ are determined 
from the normalization conditions (\ref{normalization}): 
$a_1=\frac{8}{\pi^4}b_1^4$, $b_1=\frac{372\zeta(5)}{\pi^4}$, 
$a_2=\frac{2}{93\zeta(5)}b_2^5$, and $b_2=\frac{\pi^6}{186\zeta(5)}$.  
In Figs.~\ref{trial}(a) and \ref{trial}(b), we see that 
$P^{(\beta)}_{\rm RM}(S_{\lambda})$ 
is well approximated by $P^{(\beta)}_{\rm trial}(S_{\lambda})$ 
for large $S_{\lambda}$.  
Since \eref{trialeq} implies that $P_{\rm RM}^{(\beta)}(S_{\lambda})$ 
decays exponentially for large $S_{\lambda}$, we may conclude that 
no correlation exists between avoided crossings for large $S_{\lambda}$.  
Note that , for small $S_{\lambda}$, we have 
$P^{(1)}_{\rm trial}(S_{\lambda})\simeq 5.1 S_{\lambda}^2$ and 
$P^{(2)}_{\rm trial}(S_{\lambda})\simeq 64 S_{\lambda}^3$, which 
do not reproduce the numerical data.  

We also obtained the ACSD of the following random matrix Hamiltonian.  
\begin{equation}
 \mathcal{H}_{\rm RM0}(\Lambda) = 
\mathcal{H}_0^{\rm RM} + \Lambda \mathcal{V}^{\rm RM}.
\label{rmmodel2}
\end{equation}
We checked that the ACSD of this model is the same as that of 
$ \mathcal{H}_{\rm RM}(\Lambda)$ in Eq.~(\ref{rmmodel1}).  

%%%%%%%%%%%%%%%%%%%%%%%%%%%%%%%%%%%%%%%%%%%%%%%%%%%%%%%%%%%%%%%%%%%
\section{Coupled Rotators Model}
\label{crt}

As an example of quantum systems for GOE, we consider the coupled 
rotators model \cite{Feingold83}.  We shall show that the ACSD of this 
model is well approximated by $P_{\rm RM}^{(1)}(S_{\lambda})$.  
The Hamiltonian of the system is given by 
\begin{equation}
 \mathcal{H}_{\rm CRT}(\Lambda) = L_1^xL_2^x + \Lambda (L_1^z+L_2^z),
\label{crtmodel}
\end{equation}
where $\vecvar{L}_1$ and $\vecvar{L}_2$ are angular momentums with $l_1$ and 
$l_2$, i.e., the eigenvalues of $L_1^2$ and $L_2^2$ are 
$l_1(l_1+1)\hbar^2$ and $l_2(l_2+1)\hbar^2$, respectively.  
Since $\hbar$ is a dimensionless parameter in the present unit, 
we put $\hbar=0.15875$ so that the corresponding classical system 
becomes strongly chaotic \cite{Feingold84}.  
Noting that $\mathcal{H}_{\rm CRT}(\Lambda)$ 
changes the $z$-component of the total angular momentum 
$(J^z=L_1^z+L_2^z )$ by $0,\;\pm2\hbar$ \cite{Borgonovi98}, 
the Hilbert space is divided into two subspaces corresponding to 
even $J^z/\hbar$ and odd $J^z/\hbar$.  Here, we take the subspace in 
which $J^z/\hbar$ is even.  Furthermore, we divide this subspace 
into two more subspaces corresponding to the states symmetric and 
antisymmetric under the exchange of $L_1^z$ and $L_2^z$.  Here, 
we choose the subspace corresponding to the symmetric states.  In this 
subspace, we have 1024 levels with no degeneracies when we set 
$l_1=l_2=31$.  We have confirmed that the energy level spacing 
distribution of the coupled rotators model is well approximated by 
the eigenvalue distribution of random matrices taken from GOE in the 
whole range of $\Lambda\in[0.5,1.5]$. 

\begin{figure*}[t]
\begin{center}
\includegraphics[scale=1.0]{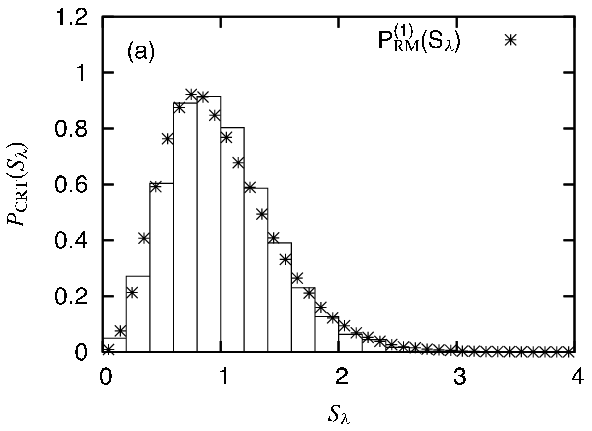}
\includegraphics[scale=1.0]{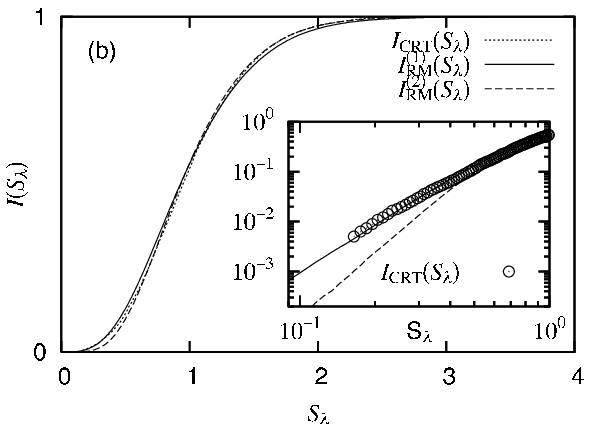}
\end{center}
\caption{(a) Histogram of $P_{\rm CRT}(S_{\lambda})$ with 
$P_{\rm RM}^{(1)}(S_{\lambda})$. 
(b) $I_{\rm CRT}(S_{\lambda})$ together with 
$I_{\rm RM}^{(1)}(S_{\lambda})$ and $I_{\rm RM}^{(2)}(S_{\lambda})$.  
The inset shows the magnified figure with a log-log plot near the origin.
}
\label{crtfig}
\end{figure*}

After the scaling (\ref{scaling}), we find 9094 avoided-crossing 
spacings.  Thus, we obtain the ACSD of the coupled rotators model 
$P_{\rm CRT}(S_{\lambda})$ and 
the cumulative distribution 
$I_{\rm CRT}(S_{\lambda})\,\left(=\int_{0}^{S_{\lambda}} 
P_{\rm CRT}(S_{\lambda}') {\rm d}S_{\lambda}'\right)$.  
In \fref{crtfig}(a), we show $P_{\rm CRT}(S_{\lambda})$ together with 
$P_{\rm RM}^{(1)}(S_{\lambda})$.  
We see that $P_{\rm CRT}(S_{\lambda})$ is roughly equal to 
$P_{\rm RM}^{(1)}(S_{\lambda})$.  
In \fref{crtfig}(b), we show $I_{\rm CRT}(S_{\lambda})$ 
with $I_{\rm RM}^{(1)}(S_{\lambda})$ and 
$I_{\rm RM}^{(2)}(S_{\lambda})$.  
The inset, which shows the magnified figure with a log-log plot, shows 
that $I_{\rm CRT}(S_{\lambda})$ follows $I_{\rm RM}^{(1)}(S_{\lambda})$.  
Therefore, we classify $P_{\rm CRT}(S_{\lambda})$ as 
$P_{\rm RM}^{(1)}(S_{\lambda})$.  

%%%%%%%%%%%%%%%%%%%%%%%%%%%%%%%%%%%%%%%%%%%%%%%%%%%%%%%%%%%%%%%%%%%
\section{Aharonov-Bohm Billiard}
\label{ab}

\begin{figure*}[t]
\begin{center}
\includegraphics[scale=2.0]{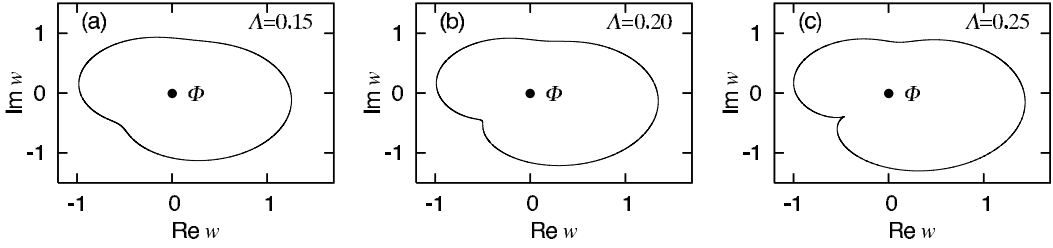}
\end{center}
\caption{Boundaries of the Aharonov-Bohm billiard with 
$\Lambda=$ (a)$0.15$, (b)$0.2$, and (c)$0.25$.}
\label{b}
\end{figure*}

Now we consider the Aharonov-Bohm billiard as an example of a quantum 
system corresponding to GUE \cite{Berry86,Robnik92,Robnik99}.  
We shall show 
that the ACSD of this system is well approximated by 
$P_{\rm RM}^{(2)}(S_{\lambda})$.  
In the billiard, a charged particle with mass $m$ and charge $q$ 
moves inside the boundary $\partial D$.  The domain $D$ is threaded by 
a magnetic flux $\Phi$ at the origin.  The Schr\"{o}dinger equation of 
this system is written as 
\begin{eqnarray}
 \frac{1}{2m}\left(-{\rm i}\hbar\nabla_{uv}-q\vecvar{A}(\vecvar{r})\right)^2
 \psi(\vecvar{r}) &=& E\psi(\vecvar{r}) \nonumber \\
 \nabla_{uv}\times\vecvar{A}(\vecvar{r}) &=& 
 \hat{\vecvar{n}}\Phi\delta(u)\delta(v), 
\label{ab_schroe}
\end{eqnarray}
where $\vecvar{r}=(u,v)$, $\psi(\vecvar{r})=0 \mbox{ on }\partial D$, 
and $\hat{\vecvar{n}}$ is the unit vector perpendicular to the billiard.  
The domain $D$ is determined by the following conformal transformation 
in the complex plane $w=u+{\rm i}v$.  
\begin{equation}
 w = z + \Lambda z^2 + \Lambda {\rm e}^{{\rm i}\frac{\pi}{3}}z^3.
\label{boundaryeq}
\end{equation}
Here, $z\;(=x+{\rm i}y)$ is a complex value in the unit disk $(|z|\le 1)$.  
The parameter $\Lambda$ determines the shape of the 
billiard.  Figures \ref{b} show the boundaries of the billiard when 
$\Lambda=0.15$, $0.2$, and $0.25$.  

Let us introduce two new parameters: 
\begin{equation}
 \alpha \equiv \frac{q\Phi}{2\pi\hbar},\qquad 
 k^2 \equiv \frac{2mE}{\hbar}.
\end{equation}
The value of $\alpha$ is restricted to the range $[0,\frac{1}{2}]$ 
from the symmetry of the system \cite{Berry86}.  We put 
$\alpha=\frac{1}{4}$, with which the corresponding classical system 
becomes most strongly chaotic \cite{Berry86}.  Now, the 
problem with getting the energy $E$ of the system in 
the Schr\"{o}dinger equation 
(\ref{ab_schroe}) results in the following eigenvalue problem 
by expanding the wave function with the Bessel function \cite{Berry86};
\begin{equation}
 \sum_{n'l'}c_{n'l'}M_{n'l'nl}(\Lambda)=\frac{c_{nl}}{k^2}.
\label{eigenprob}
\end{equation}
Here, $\{c_{nl}\}$ are components of the eigenvectors, and 
\begin{widetext}
\begin{equation}
 M_{n'l'nl}(\Lambda) 
= \frac{N_{nl}N_{n'l'}}{a_{nl}a_{n'l'}}\int_{0}^{1}{\rm d} r
 \;r \int_{0}^{2\pi}{\rm d}\phi\, {\rm e}^{{\rm i}(l-l')\phi}
 J_{|l-\alpha|}(a_{nl}r)J_{|l'-\alpha|}(a_{n'l'}r)|w'(z)|^2,
\label{ma}
\end{equation}
\end{widetext}
where $a_{nl}$ is the $n$-th zero of the Bessel function 
$J_{|l-\alpha|}(x)$ and 
\begin{equation}
 N_{nl}=\frac{1}{\sqrt{\pi}}\left|J'_{|l-\alpha|}(a_{nl})\right|^{-1}.
\end{equation}
Thus, we obtain the energy eigenvalues of the system by diagonalizing 
the Hermitian matrix $M(\Lambda)$.  

We make a $2000\times 2000$ matrix as $M(\Lambda)$, and use 
the lowest 550 levels \cite{Robnik92}.  We have confirmed that the energy 
level spacing distribution of the system is well approximated by the 
eigenvalue distribution of GUE in the whole range of 
$\Lambda\in[0.15,0.25]$.  

\begin{figure*}[t]
\begin{center}
\includegraphics[scale=1.0]{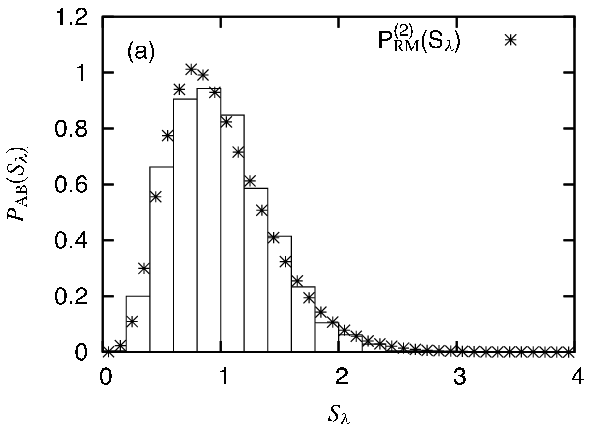}
\includegraphics[scale=1.0]{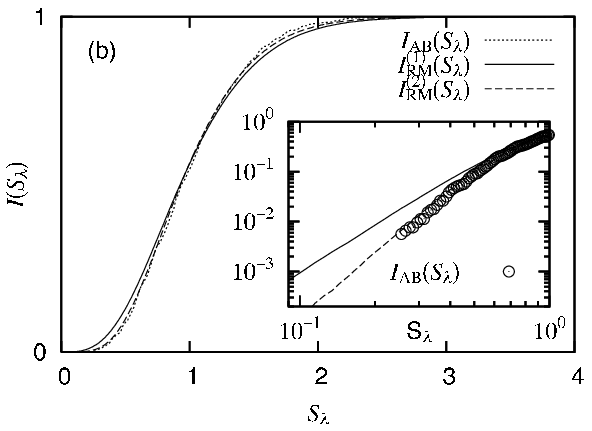}
\end{center}
\caption{
(a) Histogram of $P_{\rm AB}(S_{\lambda})$ with 
$P_{\rm RM}^{(2)}(S_{\lambda})$. 
(b) $I_{\rm AB}(S_{\lambda})$ together with $I_{\rm RM}^{(1)}(S_{\lambda})$ 
and $I_{\rm RM}^{(2)}(S_{\lambda})$.  
The inset shows the magnified figure with a log-log plot near the origin.
}
\label{abfig}
\end{figure*}

After the scaling (\ref{scaling}), we find 1050 avoided-crossing 
spacings.  Thus, we obtain 
the ACSD of the Aharonov-Bohm billiard $P_{\rm AB}(S_{\lambda})$ and the 
cumulative distribution 
$I_{\rm AB}(S_{\lambda})\,\left(=\int_{0}^{S_{\lambda}} 
P_{\rm AB}(S_{\lambda}') {\rm d}S_{\lambda}'\right)$.  
In \fref{abfig}(a), we show $P_{\rm AB}(S_{\lambda})$ together with 
$P_{\rm RM}^{(2)}(S_{\lambda})$.  
We see that $P_{\rm AB}(S_{\lambda})$ is roughly equal to 
$P_{\rm RM}^{(2)}(S_{\lambda})$.  
In \fref{abfig}(b), we show $I_{\rm AB}(S_{\lambda})$ 
with $I_{\rm RM}^{(1)}(S_{\lambda})$ and $I_{\rm RM}^{(2)}(S_{\lambda})$.  
The inset, which shows the magnified figure with a log-log plot, reveals 
that $I_{\rm AB}(S_{\lambda})$ follows $I_{\rm RM}^{(2)}(S_{\lambda})$.  
Therefore, we conclude that the ACSD of the Aharonov-Bohm billiard is 
classified as $P_{\rm RM}^{(2)}(S_{\lambda})$.  

%%%%%%%%%%%%%%%%%%%%%%%%%%%%%%%%%%%%%%%%%%%%%%%%%%%%%%%%%%%%%%%%%%%
\section{Summary} 
\label{disc}

We have studied the ACSD in quantum chaotic systems. 
The distribution $P_{\rm RM}^{(\beta)}(S_{\lambda})$ 
of the random matrix model was numerically obtained. 
We found the power law behavior for small spacings and 
the exponential decay for large spacings.  
For small spacings, we found that the powers for GOE and
GUE are $2$ and $3$, respectively. 
The exponential decay implies that the correlation 
between two avoided crossings vanishes for large spacings. 
Although these properties are established numerically,
an analytical proof for them remains to be completed.  
In \sref{crt} and \sref{ab}, we have shown by 
accurate numerical calculations that the distributions predicted by 
the random matrix model
are indeed realized in concrete quantum chaotic systems.  

The ACSD may play an important role in the quantum dynamics 
of finite fermion systems such as quantum billiards \cite{Wilkinson88}.  
It is especially worthwhile to consider the time duration until the 
system starts diffusing after the external perturbation is applied 
at zero temperature.  
As the perturbation is applied, pairs of neighboring levels in the 
energy level flow form avoided crossings.  
The quantum state at the Fermi level changes over time 
by making a nonadiabatic transition at the avoided crossing.  
The distribution of the time duration until the Fermi level 
encounters the first avoided crossing would be related to the ACSD.  
It is interesting to experimentally observe the time duration
and to understand it in the context of the ACSD.

%%%%%%%%%%%%%%%%%%%%%%%%%%%%%%%%%%%%%%%%%%%%%%%%%%%%%%%%%%%%%%%%%%%
\begin{acknowledgments}
The authors thank Professor Seiji Miyashita for his continuous 
encouragement.  One of the authors (M.~M.) also thanks 
Professor Marko Robnik for his valuable comments about 
the Aharonov-Bohm billiard.  
The simulations were partially carried out by using the computational 
facilities of the Super Computer Center at the 
Institute for Solid State Physics, the University of Tokyo.  

\noindent
$^*$Electronic address: machida@iis.u-tokyo.ac.jp\\
$^{\dag}$Electronic address: saitoh@ap.t.u-tokyo.ac.jp

\end{acknowledgments}

%%%%%%%%%%%%%%%%%%%%%%%%%%%%%%%%%%%%%%%%%%%%%%%%%%%%%%%%%%%%%%%%%%%
%\section*{References}

\end{document}